\documentclass[manuscript,times]{aastex631}

\graphicspath{figures/}
\usepackage[margins]{trackchanges}
\usepackage{savesym}
\savesymbol{tablenum}
\usepackage{graphicx}
\usepackage{xcolor}
\usepackage{siunitx}
\usepackage{amsmath}
\usepackage{tabularx}
\usepackage{graphicx}
\usepackage[figuresright]{rotating}
\usepackage{color}  % Used only by the \remark{} macro defined below.
\usepackage{subfiles} % To be able to compile included files separately
\newcommand{\au}{\ensuremath{{\textsc{au}}}}
\newcommand{\K}{\unit{K}}
\newcommand{\kms}{\ensuremath{{\rm\, km\,s^{-1}}}}

\newcommand{\forbid}[2]{[\textsc{#1}]\,#2\textrm{\AA}} % e.g. \forbid{Oi}{6300} prints [OI]6300 Angstroms
\newcommand{\NeII}{\ensuremath{\textrm{Ne}\textsc{ii}}}
\newcommand{\Oi}{\ensuremath{\textsc{Oi}}}
\newcommand{\msun}{{\rm M}_\odot}
\newcommand{\lsun}{{\rm L}_\odot}
\newcommand{\BP}{B_{\textsc{p}}}
\newcommand{\VP}{V_{\textsc{p}}}
\newcommand{\VA}{V_{\textsc{a}}}
\newcommand{\VAb}{V_{\textsc{a}0}}
\newcommand{\VK}{V_{\textsc{k}0}}
\newcommand{\RA}{R_{\textsc{a}}}
\newcommand{\cs}{c_{\rm s}}
\accepted{November 2, 2023}

\begin{document}

\title{Line profiles of forbidden emission lines and what they can tell us about protoplanetary disk winds}

\author[0000-0002-0786-7307]{Nemer. A}
\affiliation{Centre for Astrophysics and Space Science, New York University Abu Dhabi, UAE}
\affiliation{Princeton University}

\author{Goodman. J}
\affiliation{Department of Astrophysical Sciences, Princeton University, Peyton Hall, Princeton NJ 08544, USA}
%\affiliation{American Astronomical Society \\
%1667 K Street NW, Suite 800 \\



\begin{abstract}

\addeditor{AN}

Emission in forbidden lines of oxygen, neon, and other species are commonly used to trace winds from protoplanetary disks.
Using Cloudy, we calculate such emission for parametrized wind models of the magnetothermal type, following \citet{BYGY16}.  These models share characteristics with both photoevaporative and magnetocentrifugal winds, which can be regarded as end members, and are favored by recent theoretical research.  
Both broad and narrow low-velocity components of the lines can be produced with plausible wind parameters, something that traditional wind models have difficulty with.
Line luminosities, blueshifts, and widths, as well as trends of these with accretion luminosity and disk inclination, are in general accord with observations.
\end{abstract}

\keywords{Protoplanetary, disk winds, spectroscopy}


\section{Introduction}\label{sec:intro}

Protoplanetary disks (PPDs) of molecular gas and dust surround most low-mass pre-main-sequence stars (protostars).
As the term suggests, PPDs are understood to be the sites and sources of planet formation, so their properties and evolution are of great interest.
Observations show that PPDs typically disperse within a few Myr, presumably through some combination of accretion onto the star, sequestration in planets, and outflows (winds) \citep{Ercolano2017,Manara+2023,Alexander+2014}.\\

Spectroscopic observations show that PPDs are often accompanied by outflows at hundreds of \kms as well slower outflows at $\lesssim10\kms$ \citep{Hartigan+1995}.
In some of the more luminous Class~I\&II systems, the fast outflows develop into resolved, narrowly collimated jets that may extend for parsecs and terminate in Herbig-Haro objects \citep[and references therein]{Bally2016,Pascucci+2023}. Estimated mass-loss rates for both types of outflow correlate with accretion rates \citep{Hartigan+1995,Ellerbroek+2013,Rigliaco+2013,Natta+2014}.
It is generally believed that magnetohydrodynamic (MHD) mechanisms are needed to accelerate the jets, but it is not known whether these are launched very close to the star or more broadly from the PPD \citep{Frank+2014}. A disk origin seems more secure for the lower-velocity component because of the dependence of line widths on inclination, the lower blue shifts themselves, and the higher estimated mass-outflow rates \citep{Fang+2018,Simon+2016,Banzatti2019}.\\

Whether fast or slow, magnetized disk winds naturally relate outflow to accretion by extracting orbital angular momentum from the disk.
Accretion may occur via a turbulent disk viscosity independently of any wind, however, or of a magnetic connection between the disk and its wind \citep{Ercolano2017}.
In most accretion disks, the main source of turbulence is probably magnetorotational instability (MRI)\citep{Lesur+2022};
if magnetic fields are needed to explain the disk's effective viscosity, then it would seem natural that any outflow from the disk should be magnetized as well.
In PPDs, however, theoretical work suggests that MRI may be suppressed in regions likely associated with the low-velocity optical line emission ($\sim$1-20 \au) \citep{Gammie1996,Bai_2013,Turner+2014}.
Hydrodynamic mechanisms not requiring magnetic fields exist in principle, but it is not clear whether they can produce the level and type of turbulence needed to drive accretion in PPDs at observed rates \citep{Turner+2014,Fromang+Lesur2019}.\\

In any case, low-velocity disk winds almost certainly exist, so if they are not magnetized, then they must be launched by thermal pressure.
%The balance between these mass loss processes remains unclear, but the later stages of PPD evolution seem to be more rapid and hard to explain by accretion alone\citep{Gorti+2011}. 
We use the term ``photoevaporative'' for unmagnetized disk winds because in all models of such winds of which we are aware, the main source of heat is high-energy photons (FUV, EUV, X-ray) impinging on the surface of the disk and raising the sound speed to a significant fraction of the local escape velocity \citep[and references therein]{Hollenbach+1994,Gorti+2009,Gorti+2016}.
There is little doubt that photoevaporation must occur, because protostars are observed to produce these hard photons---except perhaps in the EUV, i.e. $100\unit{eV}>h\nu>13.6\unit{eV}$, which is difficult to observe because it is heavily absorbed by the ISM.
Photoevaporation can also be caused by irradiation of the outer parts of the PPD by other stars or protostars in the natal stellar association \citep[e.g.]{Adams+2004,Haworth+2016}.
Photoevaporative wind models have been successful in explaining many aspects of the observations, including line profiles and ratios \citep{Ercolano+Owen2016}.
Pure photoevaporation cannot drive accretion directly because, absent a magnetic connection, the outflow exerts no torque on the disk.
Nevertheless, a correlation between accretion and outflow could be explained if the wind were driven primarily by FUV (for which there is evidence, e.g. \citealt{Simon+2016}), since the FUV probably comes mainly from the accretion boundary layer.
Photoevaporative winds probably cannot explain the broader component (FWHM$>40\kms$) of the low-velocity emission (blueshift$<30\kms$). \citet{Simon+2016} found, from the relation between the lines' FWHM and disk inclination, that the broad component is broadened by the rotation of the disk at disk radii $\leq0.5$ AU. At those implied radii, gas at temperatures compatible with the observed lines would not be able to escape the gravitational potential, or at least not at rates sufficient to explain the line luminosity \citep{Simon+2016,Ercolano2017,Weber+2020}. \citet{Whelan+2021} and \citet{Fang+2023} performed spatially-resolved measurements of the velocity profile of optical forbidden emission lines and found the emission to be inconsistent with a thermal wind and supportive of an MHD disk wind origin of these lines.\\

In addition to providing a dynamical linkage between accretion and outflow, magnetized winds have the additional advantage in principle of being able to accelerate the outflow to a given velocity without heating it as much as would be required by a purely photoevaporative wind, hence allowing the higher-velocity gas to remain largely neutral, as suggested by some blue-shifted $[\textsc{Oi}]$ lines, and even molecular lines. However, many early magnetic models for PPD winds overdid this, explicitly or implicitly following the paradigm created by \citet[hereafter BP82]{Blandford+Payne1982}: the BP82 solutions neglect thermal pressure altogether; more importantly perhaps, their fiducial solution has a relatively large Alfv\'en radius, so that the specific angular momentum of the outflow along a given field line is asymptotically 30 times larger than the orbital angular momentum where that line meets the disk (the footpoint or wind base).  Since the angular momentum carried off by the wind balances what is lost by the disk, models based on this ``magnetocentrifugal'' paradigm tend to underpredict the ratio of the wind mass-loss rate to the accretion rate \citep{Konigl-Pudritz2000,Wardle+Konigl1993,Weber+2020}.
An additional issue is that if the magnetic field is in near equipartion with the gas and well coupled to it even at the midplane, as in some of these models, then the surface density of the disk corresponding to observed accretion rates would be far too small to form planets; this  however might allow such solutions to explain some transition disks, where substantial accretion is observed despite a drastic reduction of the surface density within a few to a few tens of \au{} \citep{Wang2019}.\\

Recent theoretical models and numerical simulations of MHD disk winds invoke weaker magnetic fields that are launched from upper layers of the disk where exposure to hard photons (FUV, EUV, \& X-rays) provides sufficient ionization, and where the gas pressure and density are much lower than at the midplane \citep{Bai+Stone2013,Gressel+2015,BYGY16,Bai2017,Bethune+2017,Wang2019,Gressel+2020}. Mass-loss rates tend to be high---comparable to accretion rates---and field lines heavily loaded, so that Alfv\'en radii ($\RA$) are only modestly larger than footpoint/launching radii ($R_0$). Magnetic and thermal pressure are at least as important as centrifugal force in launching such winds (except in some of those in \citealt{Gressel+2020}), whence they have been dubbed ``magnetothermal'' by \cite{BYGY16}. 
Magnetothermal winds can be regarded as part of a continuum of which photoevaporative and cold magnetocentrifugal winds are end members. Unlike the former, they directly connect outflow to accretion; unlike the latter, they can reconcile observed accretion rates with disk surface densities that are high enough to form planets. To date, there has been very little systematic work to explore the emission-line luminosities and profiles predicted by magnetothermal winds.\footnote{at least for the optical forbidden lines, which are the main focus of this paper; \cite{Gressel+2020} however have made interesting predictions for the $[\Oi]\,63.2\micron$ line and for millimeter-wave lines of [{\sc Ci}] and HCN based on their MHD simulations.}\\

The availability of disk-wind tracers that probe different physical conditions and  regions, along with the modelling of these tracers, makes it possible to understand the different types of disk winds and their role in PPD evolution. Forbidden emission lines from oxygen (together with other atomic species) and low ionization stages of elements such as sulfur and neon are the prime candidates for tracing disk winds, and their line ratios, if modelled correctly, can provide information about the physical properties of the emitting region. The presence of a warm ionized disk wind was first confirmed through the detection of the $[\NeII]12.8\micron$ fine structure line \citep{Pascucci+Sterzik2009,Alexander2008,Gorti+2009,Font+2004}. The line was often found to be blueshifted by a few $\kms$ relative to the stellar velocity, indicating an origin in an outflow that is approaching the observer, while the emission from the receding outflow in the other half of the disk is obstructed by the dust. The lack of high-resolution IR spectra stands in the way of collecting a large sample of PPDs to identify evolutionary trends, or the details (or components) of the line profile. The correlation observed between the neon line (luminosity, width and blueshift) and the  system variables (inclination, accretion luminosity, etc..) that were observed was successfully produced by thermal winds which led the community to accept that the [\NeII] line, in fact, traces a photoevaporative wind. However, there is a degeneracy amongst photoevaporative wind models in predicting the [\NeII] line depending on the type of radiation used to drive these winds; both X-ray and EUV driven winds can successfully predict the line, although with vastly different mass loss rates ($\sim$ 2 orders of magnitude), and the latter fails to explain the emission of other forbidden lines \citep{Ercolano+Owen2016}.\\

More progress could be made with optical forbidden lines, such as those of $\forbid{Oi}{6300}$ and $\forbid{Sii}{4068}$, since high-resolution spectra are abundant in the optical. Optical forbidden lines, e.g. $\forbid{Oi}{6300}$, have long been known to possess a low velocity component (LVC), emission blueshifted by less than 30~\kms, in addition to a high-velocity component (HVC) tracing fast (100~\kms) collimated jets (e.g., \citealt{Hartigan+1995}). EO16 simulated forbidden emission line profiles from a thermal wind that is driven by X-ray photoevaporation and found line luminosities that are in good agreement with observed LVCs. However, observational studies of high-resolution spectra by \citet{Rigliaco+2013} and \citet{Simon+2016} showed that the LVC can also be decomposed into a broad component (BC) with FWHM of at least 40 \kms and a narrower component (NC). We note that similarities between observed spectra of $\forbid{Oi}{6300}$ (composite BC and NC) and the modelled data from \cite{Ballabio+2020} (that is not fitted with 2 gaussian components) led them to conclude that gaussian fitting is not the best approach and more complex, physically-motivated model fitting (such as wind solutions) may yield greater insight than a gaussian decomposition alone. \\

From the relationship between the full width at half maximum (FWHM) of the BC and NC and the inclinations of the sample of systems observed, \citet{Simon+2016} found that the broad component likely forms in a wind that is launched at radii of $0.05$-$0.5\au$ ($0.5$-$5\au$ for the NC) if the width reflects keplerian rotation. Launching a photoevaporative wind from within $0.5\au$ requires high temperatures to achieve escape velocity. Such high temperatures will tend to ionize the gas, leaving insufficient \Oi{} to explain the observed line emission. More evidence for physically distinct emission regions for the broad and narrow components is that the [OI] 5577/6300 line ratio is typically higher in the BC than the NC \citep{Simon+2016}. \\

\cite{Pascucci+2020} detected the NC of the \Oi{} lines in spectra of both full disks (presumed to be younger PPDs) and disks depleted of dust in their inner regions (presumed to be older), albeit the emission was weaker for the latter group, which we shall refer to in this paper as ``transition disks'' (TDs).\footnote{However, it is not clear clear that TDs are always old, nor that all old PPDs pass through the TD phase \citep[and references therein]{Owen2016,Manzo-Martinez+2020,Marel+2023}. } The [\NeII] line showed an opposite relationship where the [\NeII] line is enhanced in transition disks and almost absent in full disks. This led \cite{Pascucci+2020} to conclude that the Ne line traces an outer photoevaporative wind driven by X-ray emission that can penetrate the inner regions in depleted disks while the \Oi{} BC emission comes from an inner MHD wind. The dependence of the forbidden line emission on the accretion luminosity (FUV) and the X-ray luminosity, and the inference that [\Oi] line emission comes from distinct regions created a controversy around line excitation mechanisms. Collisions with electrons and neutral hydrogen, FUV pumping, and OH dissociation have all been proposed to explain the emission \citep{Nemer_2020}.

\citet{Weber+2020} have recently computed optical forbidden-line emission for a magnetized wind based on the self-similar model of BP82.  They were able to model the [\Oi{}] lines, confirming the previous finding that the broad low-velocity component traces an inner disk MHD wind.
As explained above, however, the BP82 wind is colder and more strongly magnetized, with a larger Alfv\'en radius, than is currently thought appropriate for the low-velocity component of PPD winds; this causes \cite{Weber+2020}'s LVC to be broader, more double-peaked, and more blue-shifted than is usually observed, as they themselves point out.  \cite{Weber+2020} and others \citep{Pascucci+2020,Simon+2016} postulate that the observed emission of these lines likely traces multiple regions of the disk with different excitation and launching mechanisms, and that to reproduce these observed lines, one has to use a hybrid model that includes a combination of the above-mentioned processes. \\

In this paper we focus on the line profiles of oxygen, sulfur and neon to confirm that they can indeed trace a PPD wind, and whether those lines are able to distinguish between the different types of winds.
We construct idealized wind solutions of the magnetothermal type following \cite{BYGY16}, while varying parameters to include photoevaporative and magnetocentrifugal winds as limiting cases.
We then post-process these solutions with a radiation transport code \citep{Ferland+2017} to predict line luminosities and profiles. \\ 

\newpage
\section{Methods}\label{sec:methods}

\subsection{The wind model}\label{subsec:windmodel}

Following \cite[hereafter BYGY]{BYGY16}, our
wind solutions use a prescribed poloidal magnetic geometry: the field lines are straight, at angle $\theta'$ to the midplane, with footpoints at cylindrical coordinates $(R_0,z_0)$ on the ``surface'' of the disk.
By default, $z_0=0.15 R_0$, to approximate our expectation that the base of the wind lies several scale heights above the midplane where FUV from the star is able to penetrate and provide sufficient ionization to couple the gas to the field.
The strength of the poloidal field varies as $R^{-1}$ near the wind base, and transitions smoothly to $R^{-2}$ at $R\sim q^{-1}R_0$ along the line, $q$ being an adjustable parameter (see BYGY).
We solve the usual steady, axisymmetric, ideal-MHD equations for the flow along each field line given a prescribed density $\rho_0$ and poloidal Alfv\'en speed $\VAb=B_{\textsc{p}0}$ at the base.
The wind also has a finite sound speed $\cs$, which is taken to be constant along each field line.
Thus the flow along each line is determined by the five\footnote{As discussed in BYGY, the angular velocity $\Omega_B$ of the field line can in principle be different from the keplerian angular velocity at the wind base, $\VK/R_0$, due to a combination of vertical shear and ambipolar drift, but for simplicity we take these two angular velocities to be the same.} dimensionless input variables $(\theta',z_0/R_0,q,\VAb/\VK,\cs/\VK)$ plus three dimensionful scalings $(R_0,\rho_0,\VK)$, the last of these being the keplerian orbital velocity at the wind base, $\VK=\sqrt{GM_*/R_0}$.
After one has solved for the flow, the output variables include the positions of the critical points
\begin{equation*}
    R_{\rm slow\ magnetosonic}<R_{\textrm{Alfv\'enic}} < R_{\rm fast\ magnetosonic}
\end{equation*}
and the mass and angular momentum fluxes along the line.
Some more discussion of the fluxes is given in the Appendix.

Like \citet{BYGY16}, we call these winds ``magnetothermal'' because they are accelerated by a combination of magnetic forces (represented by $\BP$ and $\VA$) and gas pressure (represented by $\cs$), whereas the magnetocentrifugal winds of \cite{Blandford+Payne1982} are completely cold and are launched by magnetic forces alone.
(In that case, it is necessary that $z_0=0$ and $\theta'<60^\circ$.)
In the opposite limit that $\VA/\cs=0$, the wind is launched entirely by the thermal pressure and sound speed of the gas, and therefore imitates a photoevaporative wind.
We represent these unmagnetized winds by the same general scheme, with ghostly poloidal field lines that exert no force on the flow but coincide with the streamlines.
There is then only one critical point on each line---the sonic point.
While we postprocess these winds with Cloudy, as described below, the sound speed assumed in the dynamical model is not necessarily consistent with the temperature and molecular weight found by Cloudy.

\subsection{Postprocessing with Cloudy}

We have used the two-dimensional grids of emissivities from our radiation transport calculations and our grids of the wind structure to construct a three-dimensional axisymmetric volume and calculate the spectral profiles of our lines when viewed along different lines of sight (different inclinations). Based on the assumption that the disk midplane is optically thick, we omit the bottom half of the disk. The luminosities of the lines are calculated by integrating their emissivities over volume, as if each emitted line photon escapes the system. This approximation is important to our calculations, as our implementation of Cloudy is capable of radiative transfer in only one spatial dimension (the radial), so we comment here on its justification. The optical depths of the forbidden lines are small because of their small $A$ coefficients, so line scattering is neglected. 
Extinction by dust grains is included but is relatively important. 
For the fiducial model, the visual extinction along a radial ray at the base of the wind (its densest part) is a few percent. 
The hydrogen column density there is $\sim 10^{22}\unit{cm^{-2}}$, the adopted grain size $0.1\micron$, and the gas-to-dust ratio $\sim0.01\%$ by mass (i.e., depleted by a factor of a 100 compared to the ISM). \\

Our idealized wind solution does not include the whole disk, but we expect little emission to come from the regions not covered. 
We assume that the midplane is opaque to all lines of interest, including the [\NeII]~12.81\micron{} line, and consider emission from one hemisphere only; this assumption is open to question, especially for the transition disks. Also, the wind region is confined by the poloidal field lines, so for any choice of field line angle other than $90^\circ$, there will be an unmagnetized region for which the field lines would be anchored outside of the simulated region. Both of these regions are shown in blue in Fig.\ref{fig:fig1} representing a very low density region.  \\

We apply Cloudy to the density and velocity fields of our wind models, together with appropriate luminosities in energetic photons, to calculate the \forbid{Oi}{6300} and 5577\AA, \forbid{Sii}{4068}, and [\NeII] 12.81\micron{} line emission.
Our approach follows \citet{Nemer_2020}: we use the same radiation sources, chemical composition, grains, and stellar parameters. Namely, we represent the radiation source by four blackbodies with various effective temperatures. We used a blackbody with $T_{\rm eff}= 4250 \unit{K}$, $M = 0.7\msun$, and $R_* = 2.5 \unit{R_\odot}$ to represent the photospheric luminosity ($L_*=1.83\lsun$). Another with $T_{\rm eff}=12000\unit{K}$ was used to simulate the accretion radiation with a luminosity equal to the stellar luminosity (similar to the observed accretion luminosity \citep{Rigliaco+2013}). Finally we included two more blackbodies to mimic EO16's rather complex X-ray SED: one with $T_{\rm eff} = 10^6 \unit{K}$, and another with $T_{\rm eff} = 10^7\unit{K}$, dividing EO16's X-ray luminosity of $2 \times 10^{30} \unit{ergs/s}$ equally between the two. We used the same elemental abundances as reported by EO16. We also include dust with uniform grain size $0.1\micron{}$ \citep{D'Alessio1998}. We use dust-to-gas ratio $10^{-4}$ by mass in the wind \citep{Brown+2013,D'Alessio+2006,Gorti+2008} \\

However, we feed Cloudy different distributions of gas density and velocity appropriate to both photoevaporative and MHD disk winds.
We concentrate on the inner-to-intermediate regions of the disk (0.04-4 \au{}), since most of the emission in the lines listed above comes from those regions.
The calculations were performed on a spherical grid with variable cell size in the radial direction (adjusted automatically by Cloudy to optimize the calculation), and 361 points uniformly spaced in $\sin\theta\in[0,1]$. 
We integrate the line emissions over radius and over one hemisphere in $(0\le\theta\le\pi/2)$, supposing that the disk obscures the emission from the other hemisphere. Cloudy is capable only of spherical (or slab) geometries, so we use Cloudy along radial rays through the (non-spherical) wind, each such ray having its own density profile.\\  

%\begin{figure}[ht]
%centering
%\hspace*{-1in}
%\includegraphics[width=18cm,height=12cm,angle=0]{figures/fig1.pdf}
%\caption{The figure shows that gaussian fits to the line profiles can be sensitive to the choice of continuum level.  In the upper (lower) row, we pretend that the continuum covers the lower 35\% (17\%) of each \forbid{Oi}{6300} profile.  Single gaussian (\emph{green}) are fit to the upper parts of the profile (\emph{red}).  Raising the continuum generally produces a narrower gaussian.  All of these profiles are for magnetothermal$_1$ wind model.}
%\label{fig:fig1}
%\end{figure}

Cloudy produces a discrete spectrum of all the simulated lines, which we endow with line profiles suited to the local gas temperature and flow velocity.
Each cell on our poloidal grid represents a ring centered on the symmetry axis; we broaden the emission from the cell according to the gas temperature (we do not allow for turbulent broadening within cells) and assign its emission to velocity bins according to the projections of the flow onto the observer's line of sight at a specified inclination. The sum of the contributions from all such rings gives the line profile, and to account for instrumental broadening, we then convolve with a gaussian dispersion $\sigma=8\kms$ to represent the resolution of current observations. The profiles simulated in this way are usually asymmetric and skewed to the blue side, as has been found by \cite{Ballabio+2020}. Observed spectra have a varying continuum level contributed by the stellar photosphere, as well as contamination from  telluric lines. Uncertainties in the continuum level may influence the inferred line widths and (especially for asymmetric lines) blueshifts. \\ %To simulate the effect of the continuum level, we fit gaussian only to the top 65\% of the line profile (i.e., down to 35\% of the height of the peak after simulated instrumental broadening) and measure line widths and blue shifts from these fits. This is a rather crude way to account for the continuum in actual observations, but we are not attempting to fit our models to the data for any particular PPD; we merely want to prevent the extreme wings of our theoretical profiles---which would probably not be observable in real data at reasonable signal-to-noise---from unduly influencing the assigned widths and blueshifts.\\

%Clearly, however, our choice of simulated continuum level is arbitrary, and the fitted blueshifts and widths can be sensitive to that choice at the level of tens of \kms, especially for some of our more asymmetric line profiles. To demonstrate that, in Fig~\ref{fig:fig1} we chose two arbitrary levels of  continuum, and then we fit a gaussian to the part of the line that is above the continuum. As can be seen, the choice of the continuum level significantly affects the line shape, and in extreme cases could exclude one of the components; we choose a rather exaggerated misfitting case to demonstrate the effect more clearly. This casts doubt on the categorization of the lines based on their shapes as ``broad'' or "high velocity", as commonly accepted in the literature of PPDs, because it is sensitive to the procedure and can be subjective. \\
 
 Information about the wind dynamics can be deduced from the line shape as can be seen in Fig~\ref{fig:fig1}-\ref{fig:fig7}.
 The blueshift is higher at the lowest inclinations (face-on) and decreases with increasing inclination for some of our models, but in other models, the blueshift increases with inclinations up to $\sim20^o - 40^o$, but then decreases at still higher inclination. A similar trend is reported from observations for the LVC \citep{Banzatti2019} as expected for a conical wind. We fit the line profiles with a single gaussian, and then we obtain the blueshift and the FWHM from the first and second moment of the fit, respectively. We fit the gaussian to the top 99\% of the line to avoid numerical errors while fitting down to the continuum level. Those fits are all listed in Table. 2. We chose to fit with a single gaussian, contrary to the multi-gaussian fitting that is done for observations and then assigning each of the components to a distinct emission region, which shall be discussed further below. \\

\section{Results}\label{sec:results}

\subsection{Line data and comparison with observations}
 
Table \ref{tab:lumin} shows the input parameters and line luminosities for representative models of each of the three types considered in this work: magnetothermal, photoevaporative, and magnetocentrifugal.
The models are intended to be representative of the winds proposed by \cite{BYGY16}, \cite{Ercolano+Owen2016}, and \cite{Weber+2020}, respectively. 
The inner radius of the disk is set to $0.04 \au$, and the streamlines, which coincide with magnetic field lines for the magnetized winds, thread the disk at an angle $\theta'=60^\circ$ from the midplane.

The observed forbidden line luminosities range between $\sim 10^{-7} - 10^{-3} L_{\sun}$ with a median around $10^{-5}L_{\sun}$ for the \textsc{Oi} lines and $10^{-6}L_{\sun}$ for the [\NeII] line, while the $\forbid{Oi}{6300}/\forbid{Oi}{5577}$ ratio ranges from 1 to 15 with a median around 7. 
The range of blueshifts from observations is between 0 and 300 \kms for all lines; any line with blueshift more than 30 \kms is considered an HVC and below that it is an LVC. The range of FWHM is 10 to 200 \kms, with lines wider than 40 \kms considered to be BC \citep{Simon+2016,Fang+2018,Banzatti2019}.\\

\begin{deluxetable*}{c|c|c|c|c|c|c|c|c|c}
\tablenum{1}
\tablecaption{Salient properties of representative wind models. The sound speed ($\cs$) and the Alfv\'en speed ($\VA$) are normalized by the orbital speed except for the photoevaporative wind model, where $\cs$ is constant, as described in appendix A. $Z_0$ is the height of the wind base above the midplane in units of $R_0$, the cylindrical radius of the wind base. $L_1$ is the \forbid{Oi}{6300} luminosity, $L_2$ \forbid{Oi}{5577} luminosity, $L_3$ \forbid{Sii}{4068}, and $L_4$ is the [\NeII] luminosity.
\label{tab:lumin}}
\tablewidth{0pt}
\tablehead{
\colhead{Model} & \colhead{$Z_o $} &  \colhead{$c_s$} & \colhead{$v_A$} &
\colhead{$R_{in}$} & \colhead{$L_1$} & \colhead{$L_1/L_2$} & \colhead{$L_2$} & \colhead{$L_3$}& \colhead{$L_4$}\\
 &  &   & \colhead{$(v_k)$} &\colhead{$(AU)$} & \colhead{$(L_{\sun})$}& & \colhead{$(L_{\sun})$}&\colhead{$(L_{\sun})$}&\colhead{$(L_{\sun})$}\\}
%\nocolhead{Common} & \colhead{Object} &
%\multicolumn2c{Distance}

%\decimalcolnumbers
\startdata
Magneto-  & 0.15 &  0.1 & 0.1 & 0.04 & 2.3$\times10^{-5}$ & 2.4 & 9.5$\times10^{-6}$ & 2.3$\times10^{-5}$ & 5.3$\times10^{-7}$ \\ 
thermal$_1$  & $R_o$ & $(v_k)$  &  & &  &  & & &   \\
\hline
Magneto-  & 0.15 &  0.1 & 0.1 & 0.04 & 3.4$\times10^{-5}$ & 2.5 & 1.3$\times10^{-5}$ & 2.0$\times10^{-5}$ & 1.3$\times10^{-7}$ \\ 
thermal$_2$  & $R_o$ & $(v_k)$  &  & &  &  & & &  \\ 
\hline
Photo-  & 0.25 &  5.8 & 0 & 0.04 &  1.9$\times10^{-5}$ & 5.6 & 3.45$\times10^{-6}$ & 1.5$\times10^{-5}$ & 4.8$\times10^{-6}$ \\ 
evaporative  & $R_o$ & $\kms$ &  &  &  &  & & &  \\ 
\hline
Magneto- & 0 &  0.05 & 1 & 0.04 & 4.2$\times10^{-5}$ & 2.8 & 1.5$\times10^{-5}$ & 3.6$\times10^{-5}$ & 4.3$\times10^{-7}$\\ 
centrifugal$_1$ &  & $(v_k)$  &  &  &  &  & & & \\ 
\hline
Magneto- & 0 &  0.05 & 1 & 0.04 & 3.6$\times10^{-5}$ & 2.8 & 1.2$\times10^{-5}$ & 2.3$\times10^{-5}$ & 1.0$\times10^{-7}$\\ 
centrifugal$_2$ &  & $(v_k)$  &  &  &  &  & & &  \\ 
\enddata
\end{deluxetable*}

\begin{figure}[ht]
\centering
\includegraphics[width=18cm,height=12cm,angle=0]{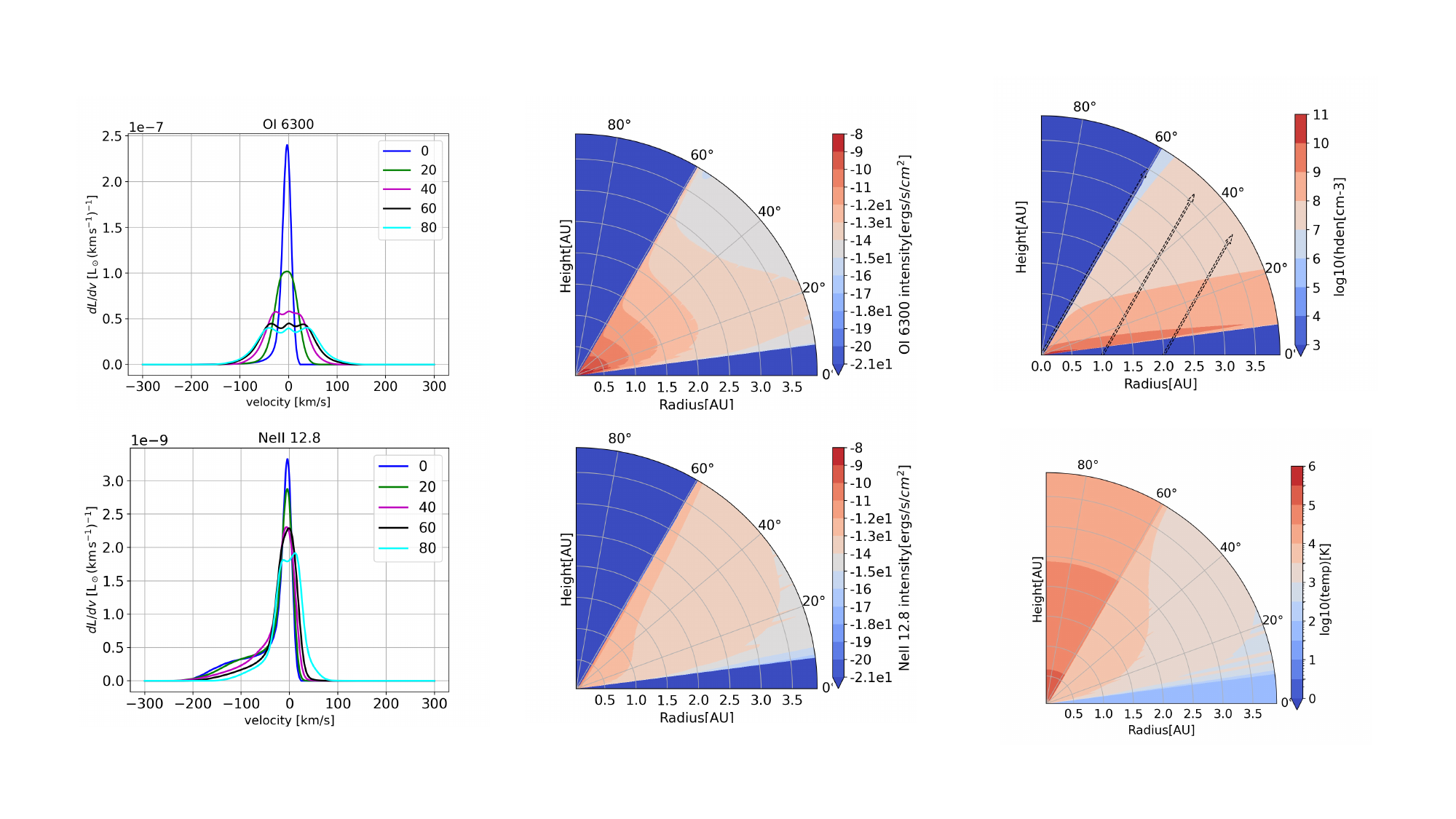}
\caption{Polar plots of emissivity for \forbid{Oi}{6300} and [\NeII] 12.81 $\micron{}$m along with their line profiles, temperature and density for our fiducial model (magnetothermal$_1$ in Table~\ref{tab:lumin})}
\label{fig:fig1}
\end{figure}

\begin{figure}[ht]
\centering
\includegraphics[width=18cm,height=12cm,angle=0]{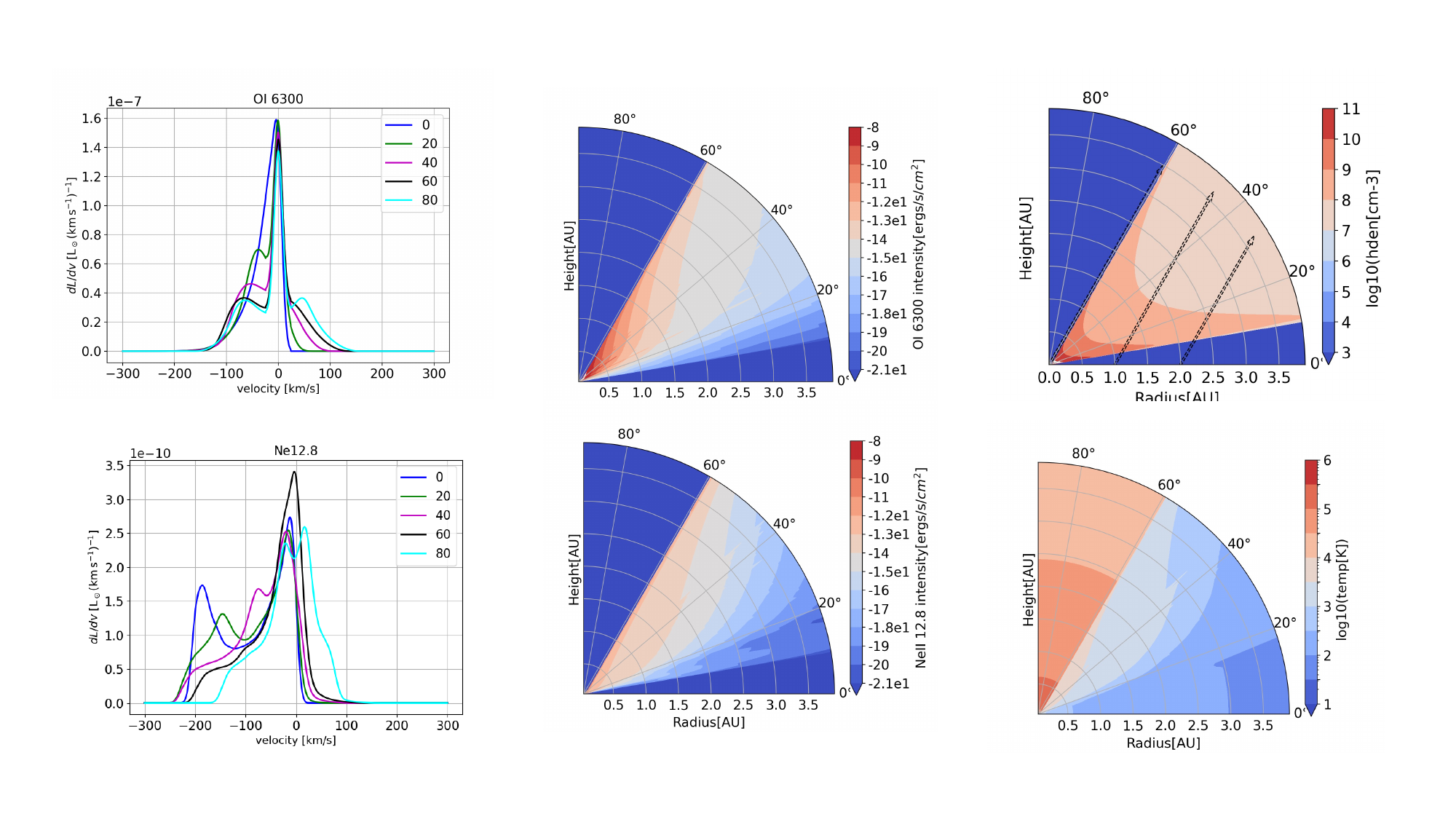}
\caption{Polar plots of emissivity for \forbid{Oi}{6300} and [\NeII] 12.81 $\micron{}$ along with their line profiles, temperature and density for our fiducial model with a different scaling to the base density, (magnetothermal$_2$ in Table~\ref{tab:lumin}).}
\label{fig:fig2}
\end{figure}

\begin{figure}[ht]
\centering
%\hspace*{-1.5in}
\includegraphics[width=18cm,height=12cm,angle=0]{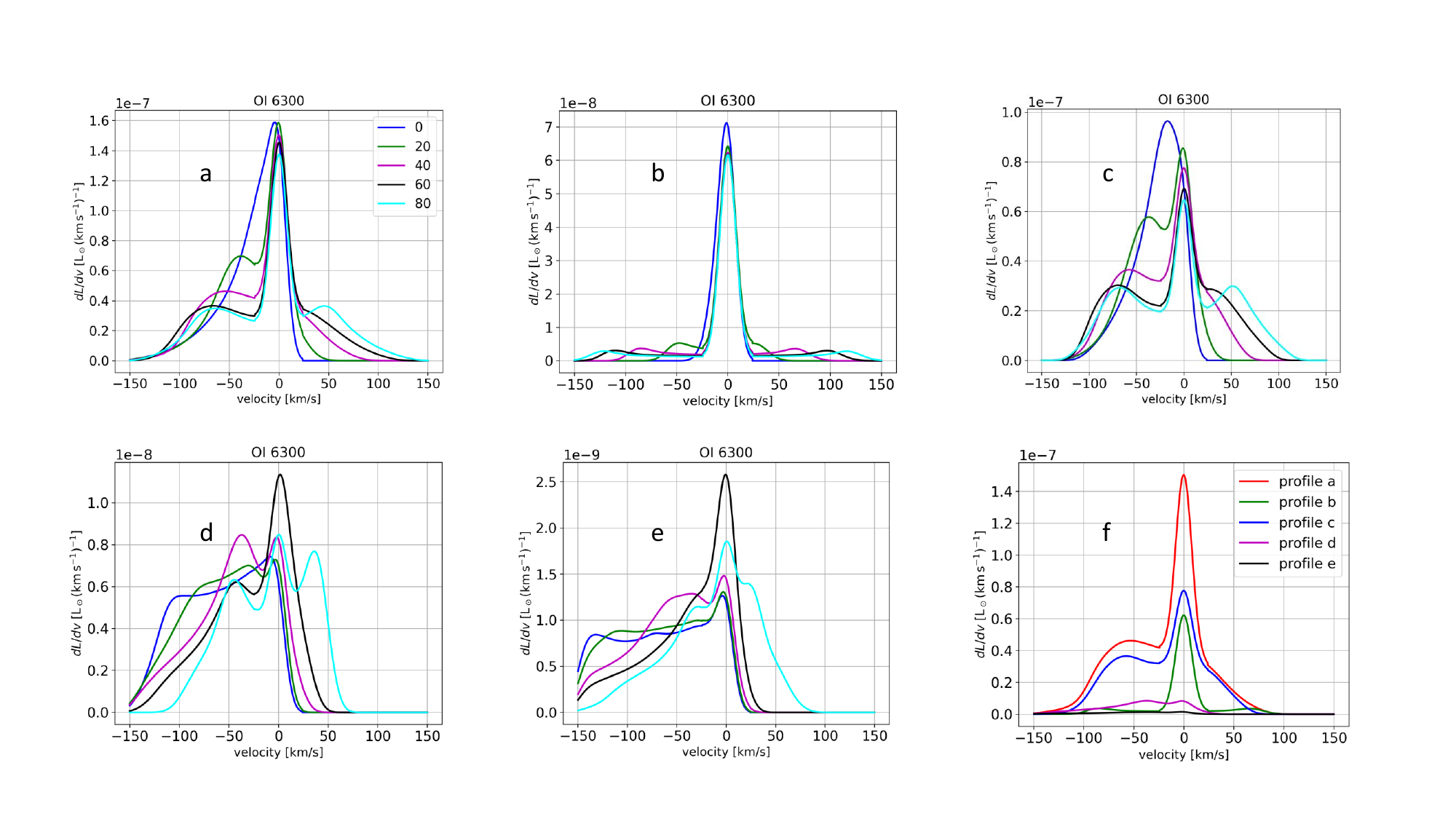}
\caption{Sampling of line profiles from different radial regions in the fiducial model at various inclinations. \emph{(a)} The full line; \emph{(b)} emission from $0.04<r<0.1\au$;  \emph{(c)} $0.1<r<0.5\au$; \emph{(d)} $0.5<r<1\au$; \emph{(e)} $1<r<4$. \emph{(f)} we show for one inclination (i=40) all the radial profiles in one panel}
\label{fig:fig3}.
\end{figure}

\begin{figure}[ht]
\centering
\hspace*{-0.75in}
\includegraphics[width=22cm,height=8cm,angle=0,trim={0 5cm 0 5cm},clip]{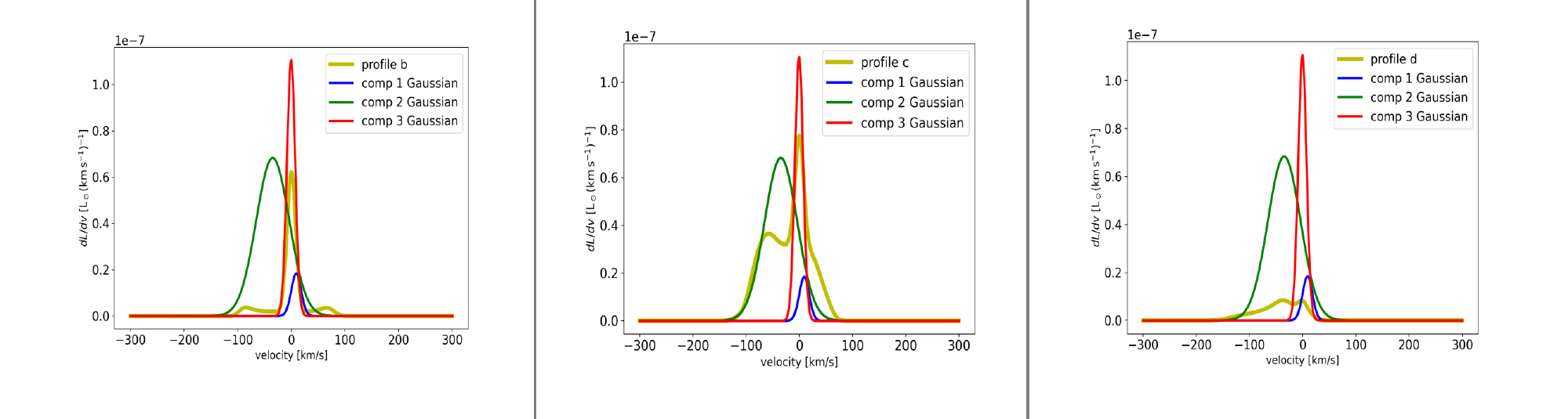}
\caption{Comparing individual components of the multi-gaussian fitting for the full profile in Fig.\ref{fig:fig3} against the emission profiles of distinct regions for the magnetothermal$_2$ model for i=20 (i.e. Fig.3b,3c,3d)}
\label{fig:fig4}.
\end{figure}

\subsection{Magnetothermal wind}
%\subsubsection{fiducial model: $magnetothermal_1$}
We choose the input parameters for our fiducial model (``magnetothermal$_1$'' in Table~\ref{tab:lumin}) to include the effects of both magnetic and thermal forces, and we optimize this input to achieve realistic mass loss and accretion rates. The details of our choices are summarized in the Appendix. 
Here we look at the spatial distribution of the line emission in our fiducial model. As can be seen from Fig.~\ref{fig:fig1}, the emission in the optical forbidden lines is concentrated in the inner regions ($<1\au$) and low latitudes ($10-30^\circ$), whereas the [\NeII] emission is more extended and mostly from higher latitudes. There are multiple reasons for this disparity. We find that each line is most luminous in regions where the density is close to its critical density, as found previously by \citet{Ballabio+2020}.
The line with the highest critical density (\forbid{Oi}{5577}) comes from the innermost regions of our fiducial model,  while [\NeII]~$12.81\micron$, which has the lowest critical density, comes from more extended regions.  
Gas temperature also has an effect.
The relatively small excitation energy of the fine-structure $[\NeII]$ line ($\approx 0.1\unit{eV}$) prefers lower temperatures ($\sim1000\K$) than the optical lines ($\sim8000\K$). 
As can be seen from Fig.~\ref{fig:fig1}, the inner regions are hotter than the outer regions.
A third factor is the penetration of the $\sim 1\unit{keV}$ X-rays needed to form [\NeII], as discussed below.\\ 

The simulated line luminosities shown in Table~\ref{tab:lumin} are generally within the range observed. We find that our predicted line luminosities are sensitive to the FUV continuum illuminating the wind, as is also observed (\S\ref{sec:intro}).  The FUV is the main heat source for our model winds, and collisional excitation of the forbidden lines depends on the gas temperature, as discussed above. The FUV heating comes mainly from the grain photoelectric effect, gas-grain collisions, H- photoionization, and photoionization of excited hydrogenic species. In addition, the oxygen lines can be excited by FUV pumping \citep{Nemer_2020}. The line luminosities for the fiducial magnetothermal wind model are broadly consistent with the observations \citep{Pascucci+2020}, and as expected, the line emission positively correlates with the FUV luminosity for the optical lines. This is expected, because the optical lines are excited by collisions, to which FUV contributes by heating the gas, and by FUV pumping. On the other hand, [\NeII] line shows a negative correlation with the FUV flux in our models. This has not been reported by observers. The [\NeII] line has a lower excitation energy, so the increase in temperature due to the increase in FUV flux tends decrease emission in this line \citep{Wan_2019}. As is well known \citep[e.g.][]{DraineBook}, the rate of excitation of ions by electron collisions tends to scale as $T^{-1/2}$, whereas the excitation of neutrals is exponentially suppressed if $kT_e$ is less than the energy of the transition, which is $\sim 2\unit{eV}$ for the $\Oi$ lines.
%For temperatures typical of PPDs (0.43 - 0.86eV) the increase in temperature will shift the maxwellian distribution of free electrons in the plasma towards higher energies leaving less electrons to excite the [\NeII] transition with energy 0.1eV. The opposite is true of the [OI] transitions with energy 2eV
\\

Line ratios can be used to constrain physical conditions in the emitting regions, provided that one understands how the upper states of the line transitions are populated.
It was reported in EO16 that the observed \forbid{Oi}{6300}/(\forbid{Oi}{5577}) ratios are mostly in the range of 1-15 with a median of about 5. 
Some authors (EO16, and others) assume that the lines are purely collisionally excited, so that the line ratio should depend only on the density and the temperature of the emitting region.
On the other hand, \citet{Gorti+2011} argue that to produce the observed line ratios by collisional excitation requires either very high densities or high temperatures ($T_e > 30,000\K$). 
They suggest that the line ratios could more naturally be produced by OH dissociation, but \cite{Fang+2018} argue against this by comparison with \forbid{Sii}{4068}. 
\cite{Nemer_2020} point out the importance of FUV pumping for the oxygen lines, an effect that is included in our present calculations.  
In short, caution is needed when using these line ratios to diagnose physical conditions in the gas.\\  

The line profiles in Fig.~\ref{fig:fig1} show a single peak with a modest blueshift and both broad and narrow width, hence would be classified as NC or BC depending on the veiwing angle.
The \forbid{Oi}{6300} comes mostly from lower latitudes and intermediate radii,  as suggested by the line shape. 
On the other hand, the [\NeII] line's emission is more extended and samples higher latitudes in the wind where the poloidal velocities are larger. As a result, the [\NeII] line has a clear blue tail and higher mean blueshift; if more emission comes from those layers, the blue tail could turn into the BC. The blueshifts are $0-3\kms$ for all the lines. \\
 
\subsubsection{Alternate scaling for the wind-base density}
Here we present a variant of the magnetothermal wind model
in which the density at the base scales as $R^{-5/2}$, more steeply than in the fiducial model ($\sim R^{-1.33}$).
As noted in the Appendix, the steeper density scaling implies a mass-loss rate $d\dot M_{\rm wind}/d\ln R_0 \propto R_0^{-1}$, and an accretion rate through the disk with the same scaling if the wind torque is the main driver of accretion. 
Such a disk and wind could not be in steady state but would develop a central hole.
This might be an mechanism for producing transition disks, but our main purpose here is to explore how a more radially concentrated wind affects the line profiles.

We can see from Fig. \ref{fig:fig2} that the density has a different structure; notably, the gas is more concentrated toward higher latitudes, where the field lines emanate from footpoints at small radii.
The line profiles in Fig.~\ref{fig:fig2} display distinct peaks that could be interpreted as broad and narrow components.
\textbf{To the best of our knowledge, this is the first physical wind model to display both a BC and a NC.}
As shown below, these components come from different regions in the wind. 
The blueshifts are $7-30\kms$ for the optical lines and $10-80 \kms$ for the [\NeII] line; the latter is at the upper limit of the LVC classification in observations \citep{Pascucci+2020}. 
This range of blueshifts in the [\NeII] is achieved by a single wind model (magnetothermal$_2$) at various inclinations and is subject to the gaussian fitting of the line in Fig.\ref{fig:fig2}.
By varying the input parameters for these magnetothermal models, we can obtain higher or lower blueshifts, depending on the balance between magnetic and thermal forces.  We find that the density at the wind base and the FUV and X-ray radiation sources correlate positively with the blueshift, but the effect is small: a few \kms for an order-of-magnitude increase in the radiation source or doubling of the base density. \\
 
The blueshift is most sensitive to the assumed Alf\'ven velocity at the wind base. When increasing the Alf\'ven velocity by an order of magnitude from the fiducial value (as an extreme following BYGY), we find that the blueshift increases by an order of magnitude from $\sim 15 - 93 \kms$ which is a significant change and can affect the classification of the line. The blueshifts and line profiles are less sensitive to the sound speed specified in our wind solutions, as long as the latter is reasonably consistent with the temperatures at which the lines of interest are strong: 3-10\kms. 
(We choose our wind parameters with this in mind, but we have not attempted an iterative reconciliation of the dynamical sound speed with the temperatures computed by Cloudy.  This might be achievable but would complicate our wind models considerably, as they currently take the sound speed to be constant along each field line.) \\

To investigate the origins of the two components seen in \forbid{Oi}{6300},  we show in Fig.~\ref{fig:fig3} the contributions to the line profile from restricted ranges of radii within the wind.  
Evidently, the broader peak is produced mainly within $0.5\au$.
As we include larger radii, we start to see a second narrower and less blueshifted peak emerging.
These findings are consistent with \cite{Simon+2016}, who found that the BC traces a gas from 0.05 - 0.5 AU, while the NC was found to trace the wind beyond 0.5 AU. \\

Furthermore, we fit multiple gaussians to the full profile at $20^o$ inclination, for which we find that three gaussian components give the best fit. In Fig.~\ref{fig:fig4} we show these three components  plotted against the emission profile from the regions described in Fig.~\ref{fig:fig3} (i.e. Fig. 3b, 3c, and 3d). One might hope that the decomposition of the line profile into gaussian components would represent distinct emission regions in a one-to-one fashion.
But as can be seen from Fig.~\ref{fig:fig4}, this is not the case. In fact, the issue that stands out the most is that the wing of any of the regional profiles is not well captured by any single component of the the 3 gaussian fit.

\begin{figure}[ht]
\centering
%\hspace*{-1.5in}
\includegraphics[width=18cm,height=12cm,angle=0]{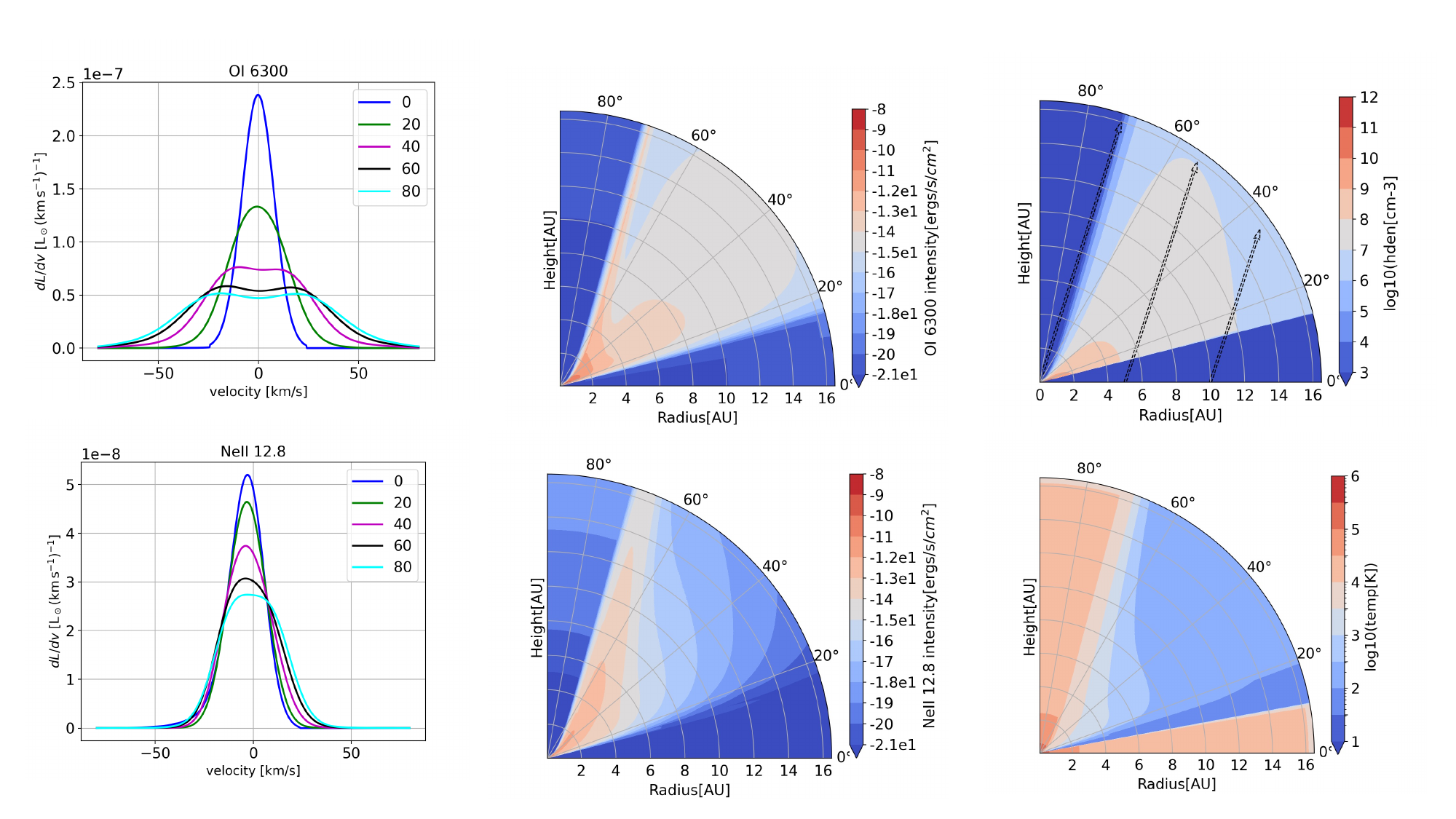}
\caption{Polar plots of emissivity for \forbid{Oi}{6300} and [\NeII] 12.81 $\micron{}$ along with their line profiles, temperature and density for photoevaporative wind model. Note that the velocity scale is smaller than for the magnetized models. The sonic point (from the wind base) for a streamline anchored at $0.5\au = 5.7\au$, at $5\au = 1.6\au$, and at $10\au = 1.4\au$ well beyond the \forbid{Oi}{6300} emission region but captures some of the [\NeII] emission as seen in the line profile.}
\label{fig:fig5}
\end{figure}

\begin{figure}[ht]
\centering
%\hspace*{-1.5in}
\includegraphics[width=18cm,height=12cm,angle=0]{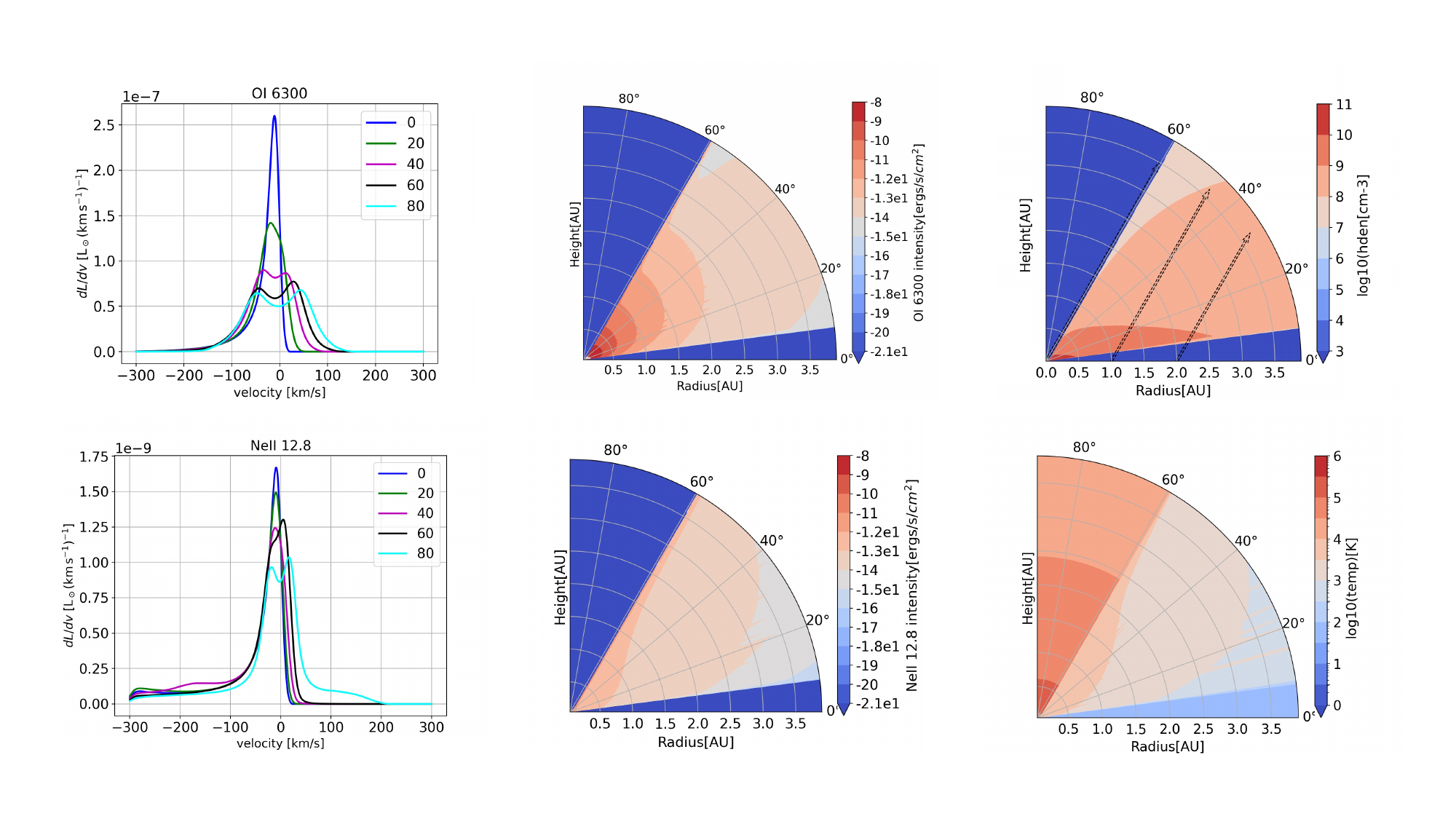}
\caption{Polar plots of emissivity for \forbid{Oi}{6300} and [\NeII]\,12.81\micron{} along with their line profiles, temperature and density for the magnetocentrifugal wind model (magnetocentrifugal$_1$ in Table~\ref{tab:lumin}).}
\label{fig:fig6} 
\end{figure}

\begin{figure}[ht]
\centering
%\hspace*{-1.5in}
\includegraphics[width=18cm,height=12cm,angle=0]{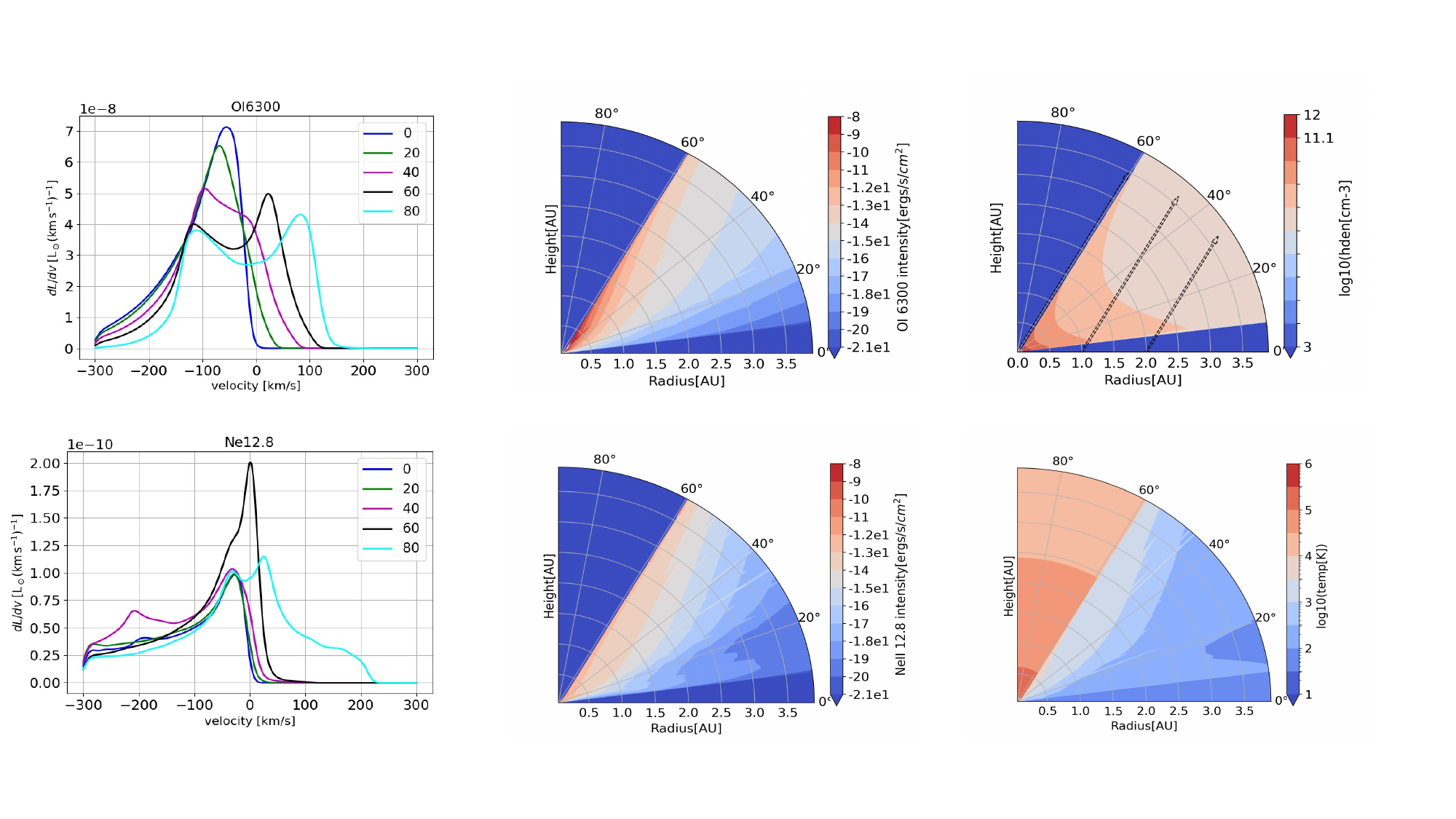}
\caption{Polar plots of emissivity for \forbid{Oi}{6300} and [\NeII]\,12.81\micron{} along with their line profiles, temperature and density for the magnetocentrifugal wind model with a different base density scaling (magnetocentrifugal$_2$ in Table~\ref{tab:lumin}).}
\label{fig:fig7} 
\end{figure}

\newpage
\subsection{Photoevaporative wind}

We construct unmagnetized wind models as described in \S\ref{subsec:windmodel}  for comparison with the photoevaporative winds of \citet{Ercolano+Owen2016,Ballabio+2020,Picogna+2019}.
As described in the Appendix, we fix the sound speed at $5.8\unit{km\,s^{-1}}$ for all radii. Note that the parameters of this model are different from those of the other four, in order to better resemble those of the above-cited works: the streamlines make angle $75^\circ$ (vs. $60^\circ$) with the midplane, the wind base lies  at $z_0=0.25 R_0$ (vs. $0.15R_0$), and the radial extent is $17\,\au$ (vs. $4\,\au$). The sonic point for the streamlines are extended beyond the emission region of the \forbid{Oi}{6300} and are indicated for a few streamlines
The optical line luminosities are on the same order of magnitude as the fiducial model. The [\NeII] line emission, on the other hand, is an order of magnitude higher than in our fiducial model which is more consistent with observations than the fiducial model. The oxygen line ratio is consistent with the observations. \\

As can be seen from Fig.\ref{fig:fig5}, the line profiles are narrower than in our fiducial model and have negligible blueshift. 
Lacking magnetic forces, the gas is accelerated to lower velocities (i.e. smaller blueshifts) that exceed the local escape velocity only at larger radii, where rotational broadening is less.
Actually, gas escapes from every radius, but the sonic point lies far from the wind base along streamlines launched from the inner disk, so that the density of the escaping gas is exponentially small and contributes negligibly to line emission.
The [\NeII] line is narrower than the oxygen lines at high inclination, suggesting that it originates at larger radii, and shows a distinct blueshift. 
In these photoevaporative models the \forbid{Oi}{6300} emission comes mostly from an almost hydrostatic atmosphere above the disk rather than a full-fledged wind, i.e. from below the sonic point.\\

\subsection{Magnetocentrifugal wind}
%\subsubsection{$magnetocentrifugal_1$}
Since the input parameters for such an MHD wind are not well constrained, we have chosen values inspired by the fiducial model of BP82, as did \cite{Weber+2020} [model ``magnetocentrifugal$_1$" in Table~\ref{tab:lumin}].
The sound speed is half of that of the fiducial model (BP82 neglected thermal pressure altogether), the Alf\'ven speed is large in order to produce a large Alfv\'en radius ($\RA/R_0=\sqrt{30}$ in BP82), and the wind is anchored at the mid plane (as for a cold and thin disk); We omit the emission from below the "surface" of the disk as done by \cite{Weber+2020}.\\

These choices produce a faster and (in post-processing) somewhat cooler wind than our fiducial magnetothermal model. The line luminosities are on the same order of magnitude as in  the fiducial model. Comparison of the density profile between the magnetothermal and the magnetocentrifugal models suggests that the densities are similar. Actually, we should have decreased the density at the wind base in inverse proportion to $(\RA/R_0)^2$ so that the fiducial and magnetocentrifugal winds exerted the same torque on the disk, and hence made for the same accretion rate. This would give a lower wind density and fainter lines (by a factor of 4). b
But we chose the current density normalization (and slightly higher Alfv\'en speed) to produce similar accretion and mass loss rates as \cite{Weber+2020} for better comparison. \\

We find that the oxygen emission comes from the surface of the disk where the gas is not significantly accelerated, similar to our findings for the magnetothermal wind model (Fig.\ref{fig:fig6}). Here the \Oi{6300} is centered at the stellar velocity and represents the NC, while the [\NeII] comes from higher layers and exhibits a blue tail.\\
\subsubsection{Alternate density scaling}
As with the magnetothermal models, here we explore the effects of as steeper wind-base density (scaling as $R^{-5/2}$ rather than as $R^{-1.33}$ as for constant $\Sigma_{\rm FUV}$).
With this steeper density scaling (model ``magnetocentrifugal$_2$" in Table~\ref{tab:lumin}), the strong magnetization and large Alfv\'en radii of the magnetocentrifugal model produce excessive blueshifts and strongly asymmetric line profiles from the higher latitudes, as can be seen from Fig. \ref{fig:fig7}. 
As a result, those lines have multiple gaussian components, which would likely be interpreted as HVC +BC, although the HVC is often attributed to a jet rather than a disk wind.
The BC component could be qualitatively consistent with the observations of \citet{Fang+2018}, who concluded that this represents a slow disk wind.
The highly blueshifted lines of the present model (magnetocentrifugal$_2$) are in agreement with \cite{Weber+2020}, but the luminosities are an order of magnitude higher than their models, as discussed above. 

\subsection{X-rays and Ne II}
The [Ne \textsc{ii}]$\,12.8\micron$ line has been postulated to be a unique signature of X-ray-irradiated protoplanetary disks \citep{Glassgold+2007,Alexander2008}, and subsequent observations of this line seem to favor an X-ray-driven photoevaporative wind \cite{Pascucci+2020}.
In this work, we find that the $\NeII\,12.8\micron$ line traces an outer (up to 20-15\au) disk wind and is strongly correlated with the X-ray radiation source assumed. In particular, the [\NeII] line traces an upper wind layer that has sufficient temperatures ($\sim 1000\K$) to excite the fine-structure level. \\

As explained in \cite{Glassgold+2007}, most of the singly ionized Neon is formed through X-ray K and L-shell ionization by photons with energies $\sim 1\unit{keV}$, followed by charge exchange with neutral hydrogen.
There are a few calculations done for the rate coefficient of collisional fine-structure excitation of the NeII line; we have consulted the most recent data from \citep{Wan_2019}, where we see that the collision strength decreases with increasing temperature, though rather slowly.
We created a planar-slab model using Cloudy to investigate the emission process for [\NeII] line. We found that the emission is proportional to the X-ray and FUV sources and density (although it saturates for higher densities) while it decreases for temperatures above $1000\K$. This confirms that the [\NeII] line emission is collisionally excited and highly dependant on the ionization of the local environment. \\

\newpage

\section{Discussion}\label{section:discussion}

This work studies the emission details of the forbidden lines based on the magnetized wind models of \cite{BYGY16}. Our aim is to demonstrate that the LVC (both BC and NC) can be produced by magnetized wind models. Current direct numerical simulations of magnetized winds rarely cover a wide enough radial range to exhibit both the BC and the NC.
\cite{Weber+2020} attempted to model forbidden lines using  the self-similar magnetocentrifugal wind of \cite{Blandford+Payne1982}; they were able to reproduce the line width and blueshift of the BC but not the line luminosities, nor the NC. We believe that this is because the magnetocentrifugal BP82 solutions, although pioneering and conceptually fundamental, are qualitatively different from the more weakly magnetized and heavily loaded magnetothermal winds that are characteristic of recent theoretical PPD work. \\

Photoevaporative wind models were shown to be successful in reproducing the observed NC of the LVC, leading \cite{Ercolano+Owen2016} and others to favor purely thermally driven winds. Even the most recent photoevaporative wind models fail to reproduce the BC of the LVC, however, and have some discrepancies with the observed line. \citet{Weber+2020} models suffer from much lower luminosities than observed, and double-peaked emission for the BC. \citet{Ballabio+2020} models cannot reproduce the observed line widths and blueshifts simultaneously, and they could only model NLVC but not the BLVC. \\

The [\NeII] line has been among the earliest spectroscopic evidence for disk winds from PPDs, and the formation of this line requires strong radiation sources (EUV or X-ray) typically associated with photoevaporative disk winds \citep{Ercolano+Owen2010}.
\cite{Pascucci+2020} recently conducted a survey of known PPD emitters of the [\NeII] line with highest resolution available for these objects, and they found that the line prevails more strongly in transition disks (TD), which (as the name suggests) may represent a later phase of PPD evolution \citep[but see][]{Owen2016}. They believe that this correlation is mainly due to the key role that the X-ray radiation plays in creating enough ionized neon to observe this line emission; the dependence of the [\NeII] line on the X-ray source in PPDs was also suggested in the theoretical work of \cite{Glassgold+2007} before such a large sample of observations was available. Hence it is believed that the X-ray radiation is less absorbed in the inner regions for disks with inner holes allowing it to reach the upper layer of the wind to excite the NeII line (where the temperature is appropriate for emission).\\

We wanted to compare line profiles from our fiducial model with other magnetized wind models. \cite{Weber+2020} performed the most recent of these simulations, where they relied on the analytic description of an MHD wind based on the work of \citep{Blandford+Payne1982} axisymmetric self-similar solutions for a magnetocentrifugally driven wind. 
The problem with these ``cold'' magnetocentrifugal wind models is that they are too strongly magnetized and too lightly mass-loaded, with the result that the velocities are too large, leading to excessive blueshifts and highly skewed profiles.  \\

Our magnetothermal model based on the work of \cite{BYGY16} is successful in reproducing both the BC and NC of LVC of the [OI] lines, and the LVC of the [\NeII] line; the latter usually lacks sufficient observational resolution to be decomposed into BC and NC \citep{Pascucci+2020}. Our fiducial model represents only one of many plausible choices for the input parameters.
The predicted line shapes depend on the parameters of the radiation sources, the sound speed of the wind, but are most sensitive to the assumed magnetic field and the density at the wind base. 
We confirm that the poloidal velocity component is mainly responsible for the blueshift, whereas the width of the line is determined by the local keplerian velocity, which depends on the distance of the emission region from the star. 
This is clear from the magnetized models, where we see broader, more blueshifted emission that comes from a wind launched in the inner regions accelerated predominantly by the magnetic field. 
Conversely, the photoevaporative model displays narrower, less blueshifted lines because the emission comes from larger radii where the rotational broadening is less.\\

\begin{figure}[ht]
\centering
%\hspace*{-1.5in}
\includegraphics[width=18cm,height=12cm,angle=0]{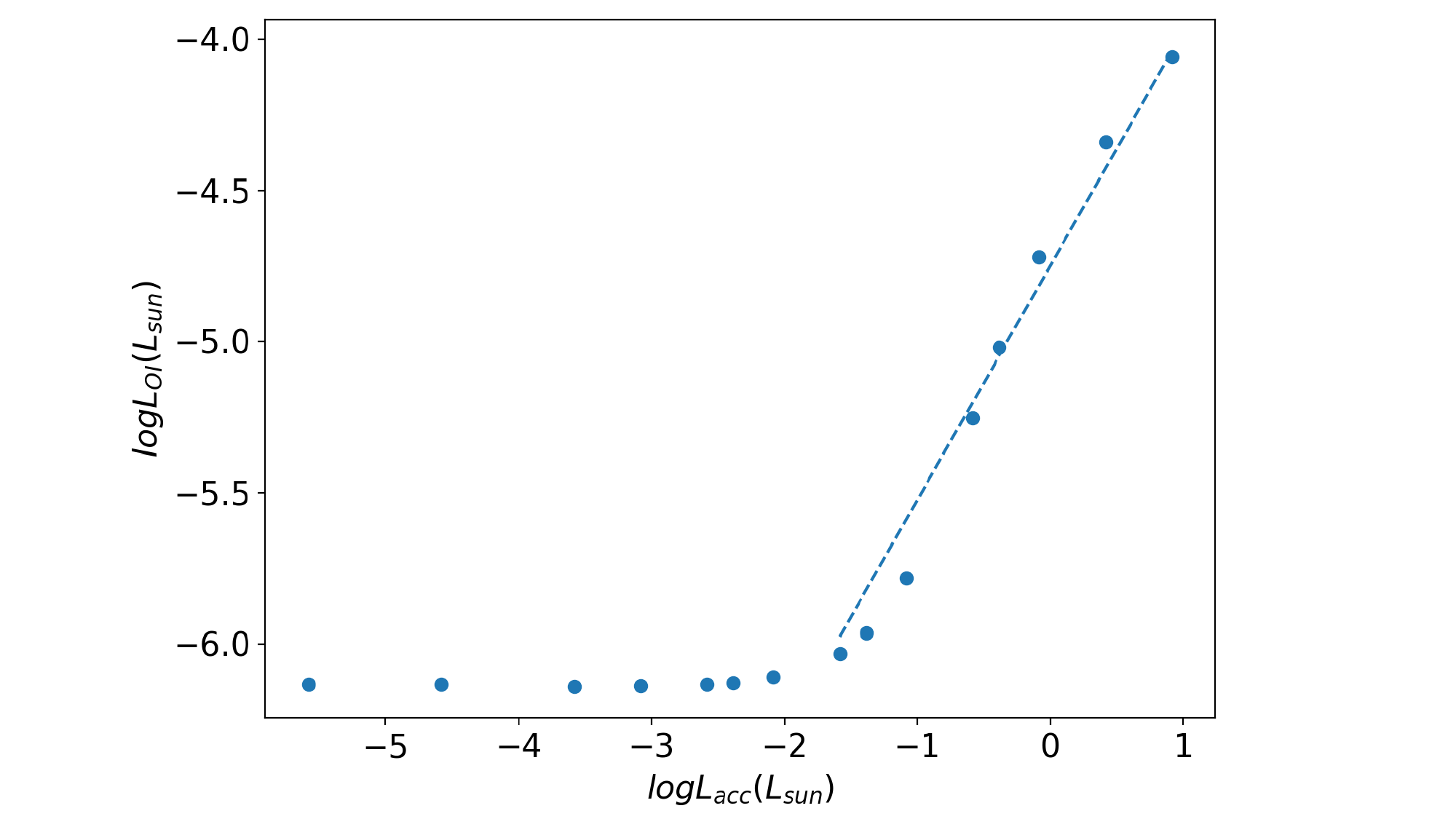}
\caption{[OI] 6300 luminosity vs accretion luminosity relation based on the simulation results of our fiducial model (orange). We overlay a fit to the data points (dashed) which is explained further in the text.}
\label{fig:fig8} 
\end{figure}

Solar mass T-Tauri stars have a wide range of observed accretion luminosities, but the majority of these systems have $L_{acc} \sim 0.01 - 0.1$ $L_{sol}$ \citep{Manara+2023,Fang+2018}. Usually, the accretion luminosity is not measured directly, because the source is mostly in the UV and is efficiently absorbed before reaching the observer, but inferred from tight correlations with H lines \citep{Fang+2018}. \citet{France+2023} directly measures the UV $H_2$, $Ly_{\alpha}$, C IV $1600 \AA$ bump, and FUV continuum between $912 - 1760 \AA$ (the major constituents of the accretion FUV continuum) using HST observations, and confirm the results of previous studies about the typical luminosity of this source. We approximate the shape of the FUV source with a blackbody with $T_{eff}$ = 12,000K following the previous theoretical work \citep{Ercolano+Owen2016,Wang2019,Gorti+2009}. Our choice of $L_{acc} = 1.8$ $L_{sol}$, although within observational constraints, is an extreme case according to the observed sample. To better compare with observations , we show in Fig.\ref{fig:fig8} the relationship between $L_{acc}$ and $L_{OI}$ for a wide range of $L_{acc}$. We find that for log($L_{acc}/L_{sun}$) $\geq$ -2, the simulated OI luminosities are -6.2 $\leq$ log($L_{OI}/L_{sun}$) $\leq$ -4 which is within the observed range for these lines \citep{Banzatti2019,Rigliaco+2013,Fang+2018}. Though, it seems that the relationship plateaus for log($L_{acc}/L_{sun}$) $\leq$ -2 as was reported by \citet{Ercolano+Owen2016} in their independent theoretical work perhaps because there is a process that suppresses the OI excitation for lower temperatures which is not included in current models. Moreover, \citet{Fang+2018} finds the correlation to follow log $L_{OI} \sim$ 0.6 log $L_{acc} - 4.07$ similar to our model with a relation  log $L_{OI} \sim$ 0.77 log $L_{acc} - 4.74$ for log$(L_{FUV}/L_{sun}) \geq$ -2. It is evident that the FUV source has a strong connection to the OI emission, and careful treatment of this source, informed by recent observational work, should be included in future theoretical work.\\

It is commonly accepted that the [\Oi] lines can be broken into broad and narrow components (BC, NC), and that these components trace distinct regions. This conclusion relies on the multi-gaussian fitting procedure used in analyzing the observations. We can see from our models that the lines are highly asymmetric and fitting a gaussian (or multiple gaussians) can be misleading. Moreover, we saw that a simple change in the level of the continuum assumed could change the shape of the fit to the line. The inference of distinct regions based on a decomposition of the observed lines into gaussians has been challenged by many authors who point out that the gaussian fitting does not clearly separate the flux of the overlapping component in the observed spectra \citep{Banzatti2019}, and that the line is inherently asymmetric \citep{Ballabio+2020} and not purely keplerian \citep{McGinnis+2018}. The above-mentioned observational limitations could introduce subjectivity to the fitting procedure. In future, we hope to explore whether fitting, both the modelled and observed spectra, with more robust procedures such as machine-learning techniques, could substantially enhance the analysis.  \\

Our models have their limitations and could be made more self-consistent. 
The sound speed assumed in launching the flow is prescribed and constant along each field line (or streamline), and no attempt is made to reconcile it with the temperature computed by Cloudy in post-processing. As in \cite{BYGY16}, the geometry of the magnetic field is prescribed (with two adjustable parameters, $\theta'$ and $q$); the flow along each line is computed in steady-state, but force balance across lines is ignored. Ambipolar diffusion and other non-ideal MHD effects have been neglected, though these could be important for launching the wind and for heating it after it is launched. Our models already have several adjustable parameters, however, so that while it will some day be important to make more complete and realistic models for comparison with data, the present state of both observations and theory may not yet justify the effort involved.\\

\section{Conclusion}\label{sec:conclusion}

In this work, we process the analytic wind solution from \cite{BYGY16} through the radiation-transport code Cloudy to obtain simulated emission spectra of PPDs. We then modify the analytic wind solution to produce models very similar to photoevaporative and magnetocentrifugal wind solutions to compare with our fiducial magnetothermal model and discuss the discrepancies. Our aim is to provide insight into current studies of wind tracers, so we focus on common observables such as the blueshift and FWHM in relation to physical properties like the magnetic field and sound speed. \\

Our simulations suggest that magnetic forces are needed to launch a wind close to the star, where the BC of the oxygen lines forms. Hence, magnetothermal wind models are needed to explain the observed complex spectra.  Photoevaporative wind models can explain only the NC, because at temperatures compatible with the forbidden oxygen lines, the sound speed is insufficient to launch a vigorous wind from small radii.
Furthermore, at least in our photoevaporative models, much of the oxygen emission comes from a quasi-hydrostatic atmosphere above the disk, as reflected by its small blueshifts, whereas the [\NeII] better samples the disk wind.

In this work, we have imitated standard observational procedure by fitting gaussian profiles to each of our model lines and declaring the blueshift and FWHM based on these fits. However, as our computed line shapes are often far from gaussian and have extended tails that might be difficult to discern above the continuum in a real observation, the use of gaussian fits may not make the best use of the available data for diagnosing the dynamics and physical conditions of disk winds. 

\begin{deluxetable*}{c|c|c|c|c|c|c|c|c|c|c}
\tablenum{1}
\tablecaption{A table that lists the blueshifts and FWHM of the \forbid{Oi}{6300} line for different inclinations from a single gaussian fitting \label{tab:fits}.}
\tablewidth{0pt}
\tablehead{
\colhead{Model} & \colhead{$v_0$} &  \colhead{$v_{20}$} & \colhead{$v_{40}$} &
\colhead{$v_{60}$} & \colhead{$v_{80}$} & \colhead{$FWHM_0$}& \colhead{$FWHM_{20}$}&
\colhead{$FWHM_{40}$} & \colhead{$FWHM_{60}$} & \colhead{$FWHM_{80}$} \\
 &\colhead{$(km/s)$} & \colhead{$(km/s)$} & \colhead{$(km/s)$}  & \colhead{$(km/s)$} &\colhead{$(km/s)$} & \colhead{$(km/s)$} & \colhead{$(km/s)$}& \colhead{$(km/s)$}&\colhead{$(km/s)$}&\colhead{$(km/s)$}\\}
\startdata
Magneto-  & -3.8 & -4.7 & -4.8 & -3.4 & -1.2 & 23.4 & 51.0 & 87.4 & 114.5 & 128.8 \\ 
thermal$_1$ & & & & & & & & & &   \\
\hline
Magneto- & -16.1& -15.3 & -11.3 & -3.1 & -1.7 & 59.6 & 78.9 & 91.4 & 105.3 & 112.5\\ 
thermal$_2$ & & & & & & & & & &   \\ 
\hline
Photo-  & -0.19 & -0.08 & -0.08 & -0.06 & -0.02 & 18.7 & 30.4 & 50.4 & 66.4 & 75.1 \\ 
evaporative  & & & & & & & & & & \\ 
\hline
Magneto-  & -15.6 & -18.6 & -17.8 & -11.6 & -3.6 & 49.6 & 77.3 & 111.1 & 136.8 & 157.5  \\ 
centrifugal$_1$ & & & & & & & & & & \\ 
\hline
Magneto- & -93.0 & -90.5 & -79.3 & -53.4 & -19.1 & 143.9 & 150.3 & 182.6 & 237.5 & 273.8 \\ 
centrifugal$_2$ & & & & & & & & & & \\ 
\enddata
\end{deluxetable*}

\newpage

\section*{Acknowledgments}

This work was supported by NASA grant 17-ATP17-0094.

\newpage
\appendix
\section{Mass-loss and accretion rates}
The solution for the flow along each field line is found by the method described in \cite[hereafter BYGY]{BYGY16}. Each such solution is determined by specifying the following control parameters:
\begin{itemize}
    \item $R_0$: the cylindrical radius at the wind base, where the field line meets the surface of the disk;
    \item $z_0$: the height above the midplane at the wind base;
    \item $\theta'$: the angle made between the field line and the local horizontal ($z=\mbox{const}$)---this is called simply $\theta$ in BYGY;
    \item $q$: the field-line divergence parameter, which determines where the prescribed poloidal field transitions from $\BP\propto r^{-1}$ to $r^{-2}$ [see eq.~(2) of BYGY]
    \item $\rho_0$: the mass density at the wind base;
    \item $\hat V_{\textsc{a,p},0}$ (abbreviated $\hat V_{\textsc{a}0}$): the ratio of the poloidal Alfv\'en speed $\BP/\sqrt{4\pi\rho_0}$ at the wind base to the keplerian orbital velocity $\VK=\sqrt{GM_*/R_0}$;
\end{itemize}
In addition to these, one specifies the angular velocity of the field line ($\omega$).  But in the present work, we take $\omega=\VK/R_0=\sqrt{GM_*/R_0^3}$.

The solutions for the flow along a given field line are found in dimensionless variables scaled by the physical variables at the line's footpoint: $(R_0,\VK,\rho_0)$
The scaled variables here by hats: e.g., $\hat z_0=z_0/R_0$ and $\hat\VA\equiv(\BP/\sqrt{4\pi\rho})/\VK$.
After solving the equations described in BYGY for the slow and fast magnetosonic radii ($R_{\rm sms}$, $R_{\rm fms}$), Alfv\'en radius ($R_{\textsc{a}}$), specific angular momentum $\hat l=\hat\omega\hat \RA^2$, and mass-flux parameter [eq.~(3) of BYGY]
\begin{equation}\label{eq:kdef}
   k\equiv \frac{4\pi\rho\VP}{\BP}\,,\qquad \hat k\equiv \frac{k}{\sqrt{\rho_0}}\,,
\end{equation}
we have everything necessary to calculate the scaled flow variables 
$\hat R,\,\hat z,\,\hat\rho,\,\hat\VP,\,\hat v_\phi,\,\hat B_\phi$
along the line, although $\hat B_\phi$ is not needed for calculating the emission lines and their profiles.

In principle it should be possible to use the same methods to calculate an unmagnetized (i.e., photoevaporative) wind by taking the limit $\BP\to 0$.
In practice, the collapse of the three critical points $(R_{\rm sms},\RA,R_{\rm fms})$ into a single one---the sonic point, $R_{\rm s}$---creates difficulties, so we have a separate code for that case.
The poloidal streamlines have the same shape and divergence as in the magnetized case, but $\hat B_\phi=0$ so that the specific angular momentum is simply $l=Rv_\phi$ and there is no magnetic contribution to the specific energy of the flow (Bernoulli constant).
The mass-flux constant is no longer scaled by $\BP$, but can be interpreted as $k'=4\pi\rho\VP\times (B_{\textsc{p},0}/\BP)$ as $\BP\to0$, instead of \eqref{eq:kdef}.
To compare the mass-loss rates of the magnetized and unmagnetized cases more clearly, it is convenient to define a dimensionless mass flux parameter $\hat\mu$ as
%\footnote{\remark{To get $\hat\mu$ as defined here, multiply {\tt windy.py}'s $\mu$ by $\hat V_{\textsc{a}0}(4\pi)^{3/2}$; and {\tt pewind.py}'s $\mu$ by $4\pi$.}}
\begin{equation}
    \hat\mu\equiv 4\pi\frac{V_{\textsc{p}0}}{\VK} = \begin{cases} \sqrt{4\pi}\,\hat V_{\textsc{a}0}\,\hat k & \mbox{if } \BP\ne0\\
    \hat k' & \mbox{if } \BP=0.\end{cases}
\end{equation}
In terms of this parameter, the mass-loss rate of the wind per logarithmic interval in radius becomes
\begin{equation}\label{eq:Mdot}
    R_0\frac{d\dot M_{\rm wind}}{dR_0} = \hat\mu \sin\theta'\,(R_0^2\rho_0\VK)\,.
\end{equation}

\subsection{Scaling of the density at the wind base with radius}

Following BYGY (see in particular their Appendix), we adopt the minimum-mass solar nebula (hereafter MMSN) and suppose that the wind base lies at the depth (column density) into the disk to which FUV penetrate, nominally $\Sigma_{\textsc{fuv}}\approx 10^{-2}\unit{g\,cm^{-2}}$ \citep{Perez-Becker+Chiang2011}.
At 1~AU in the MMSN where the density below this level (i.e., closer to the midplane) is constant at $280\unit{K}$, the base lies at $z_0\approx 4.5 H$ off the midplane, with $H\approx 0.034 (R_0/{1\au})^{5/4}\au$ being the gaussian scale height, so that $\rho_0(1\au)\approx 5.6\times10^{-14}\unit{g\,cm^{-3}}$ [see eq.~(33) of BYGY].
However, our adopted value was $\rho_0(1\au)=1.4\times10^{-14}\unit{g\,cm^{-3}}$, corresponding to $z_0/H\approx 4.8$ rather than $4.5$.

Thus for our fiducial model with $\hat V_{\textsc{ap}0}=\hat c_{\rm s0}=0.1$, where
$\hat\mu\approx0.06931$, $d\dot M_{\rm wind}/d\ln R\approx 0.90\times10^{-8}\unit{M_\odot\,yr^{-1}}$ at $R_0=1\au$.
%The corresponding photoevaporative wind with the same dimensionless sound speed but $\hat V_{\textsc{ap}0}=0$ has $\hat\mu=2.7\times10^{-19}$ and $d\dot M/d\ln R\sim 10^{-24}\unit{M_\odot\,yr^{-1}}$.
For our photoevaporative wind, we took $c_{\rm s}=5.8\unit{km\,s^{-1}}$ at all radii,  corresponding to neutral atomic gas at $T=5000\K$ and solar abundance, so that $\hat c_{\rm s0}=0.194$ at $1\au$: in this case, $d\dot M_{\rm wind}/d\ln R\approx 1.4\times10^{-10}\unit{M_\odot\,yr^{-1}}$ at $1\au$.

To scale these results to other values of $R_0$, we took two approaches.
Our first was to take the density at the wind base to scale as $R^{-5/2}$; this was used for the models ``magnetothermal$_2$" and ``magnetocentrifugal$_2$" in Table~\ref{tab:lumin}.
This corresponds approximately to scaling the density at the wind base in proportion to the density at the midplane (in fact, $\rho_{\rm mid}\propto R^{-11/4}$ in the MMSN).
Note from eq.~\eqref{eq:Mdot} that if $\hat\mu\sin\theta'$ is independent of radius, a power-law dependence $\rho_0\propto R_0^{-\alpha}$ for the density at the wind base implies $d\ln\dot M_{\rm wind}/d\ln R_0\propto R_0^{3/2\,-\alpha}$, which reduces to $R_0^{-1}$ for $\alpha=3/2$.
For a constant value of $\RA/R_0$ and therefore of the ejection index (see below), the accretion rate $\dot M_{\rm acc}$ through the disk has the same scaling as $d\ln\dot M_{\rm wind}/d\ln R_0$.
Hence under this first prescription for $\rho_0(R)$, both the mass-loss rate of the wind and the accretion rate through the disk are strongly concentrated toward small radii.

As an alternate---and arguably more physical---approach, we identify the wind base with the surface density $\Sigma_{\textsc{fuv}}$ to which the FUV penetrates, and we take this to be independent of radius.
We then solve for $z_0$ from
\begin{equation}\label{eq:baseloc}
    \frac{\Sigma_{\textsc{mmsn}}(R_0)}{\sqrt{2\pi H(R_0)}}\int\limits_{z_0}^\infty e^{-(z/H)^2/2}\,dz = \Sigma_{\textsc{fuv}},\quad\mbox{with}\quad \Sigma_{\textsc{mmsn}}(R_0)=1700 (R_0/1\au)^{-3/2}\unit{g\,cm^{-2}}\,.
\end{equation}
We find that $z_0/H$ decreases radially outward very slowly:  from $5.08$ at $R_0=0.1\au$ to $3.56$ at $R_0=10\au$.
Over the same radial range, the density $\rho_0\equiv\rho(R_0,z_0(R_0))$ scales quite accurately as $R_0^{-1.33}$, only slightly more steeply than $H^{-1}$.
Since $\VK\propto R_0^{-1/2}$, it follows from eq.~\eqref{eq:Mdot} that $d\dot M/d\ln R\propto R_0^{0.17}$ at fixed $\hat\mu$, i.e. for a given dimensionless model; and $\dot M_{\rm acc}$ has the same scaling.
This second approach was used for all of the remaining models in Table~\ref{tab:lumin}.

It cannot be overemphasized that whereas the sound speed of a wind that produces the forbidden \Oi{} lines is rather well constrained---the temperature must be 3000-10,000~K---at the present time there are very few \emph{a priori} constraints on the magnetic field (and hence $\VAb$), except by requiring that it be strong enough to explain observed accretion rates.
Following eqs.~(31)-(34) of BYGY and the scaling $\rho_0\propto R_0^{-1.33}$ found here, this leads to 
\begin{equation*}
    \hat V_{\textsc{ap}0}\approx 0.026 \left(\frac{R_0}{1\au}\right)^{-0.085}\left(\frac{\dot M_{\rm acc}}{10^{-8}\unit{M_\odot\,yr^{-1}}}\right)^{1/2}
\end{equation*}

Alternatively, we may relate the accretion rate to the mass-loss rate of the wind via the \emph{ejection index} [eq.~(17) of BYGY]
\begin{equation}\label{eq:ejection-index}
    \xi \equiv \left.\frac{d\dot M_{\rm wind}/d\ln R}{\dot M_{\rm acc}}\right|_{R=R_0} = \frac{1}{2}\frac{1}{(\RA/R_0)^2-1}\,;
\end{equation}
this follows from angular-momentum conservation if viscous/turbulent transport within the disk is small compared to the torque exerted on the disk by the wind.
For our fiducial model (magnetothermal$_1$ in Table~\ref{tab:lumin}), $\RA= 2.795$, whence $\dot M_{\rm wind}= 0.90\times10^{-8}\unit{M_\odot\,yr^{-1}}$ corresponds to $\dot M_{\rm acc}\approx 1.2\times10^{-7}\unit{M_\odot\,yr^{-1}}$.

\end{document}